\documentclass[a4paper,11pt]{article}
\usepackage{pos}

\title{Silicon detector R$\&$D for the future Electron-Ion Collider}

\author*[a]{Xuan Li}

\affiliation[a]{Physics Division, Los Alamos National Laboratory,\\
  Los Alamos, New Mexico, United States}

\emailAdd{xuanli@lanl.gov}

\abstract{The high-luminosity high-energy Electron-Ion Collider (EIC) to be built at Brookhaven National Laboratory (BNL) will provide a clean environment to study several fundamental questions in the high energy and nuclear physics fields. A high granularity and low material budget silicon vertex and tracking detector is required to provide precise measurements of primary and displaced vertex and track reconstruction with good momentum and spatial resolutions. The reference design of the EIC silicon vertex and tracking detector includes the Monolithic Active Pixel Sensor (MAPS) based vertex and tracking subsystem and the AC-Coupled Low Gain Avalanche Diode (AC-LGAD) based outer tracker, and it has the track reconstruction capability in the   pseudorapidity region of -3.5 to 3.5 with full azimuthal coverage. Further detector geometry optimization with a new magnet based on the EIC project detector reference design are being performed by the newly formed ePIC collaboration. The latest ePIC vertex and tracking detector geometry and its performance evaluated in simulation will be presented. Details of the EIC silicon vertex and tracking detector R$\&$D, which include the proposed detector technologies, prototype sensor characterization and detector mechanical design will be discussed as well.}

\FullConference{%
  10th International Workshop on Semiconductor Pixel Detectors for Particles and Imaging (Pixel2022)\\
  12-16 December 2022\\
  Santa Fe, New Mexico, USA}

\begin{document}
\maketitle

\section{Introduction to the Electron-Ion Collider (EIC)}
\begin{figure}[!ht]
\centering
\includegraphics[width=0.85\textwidth]{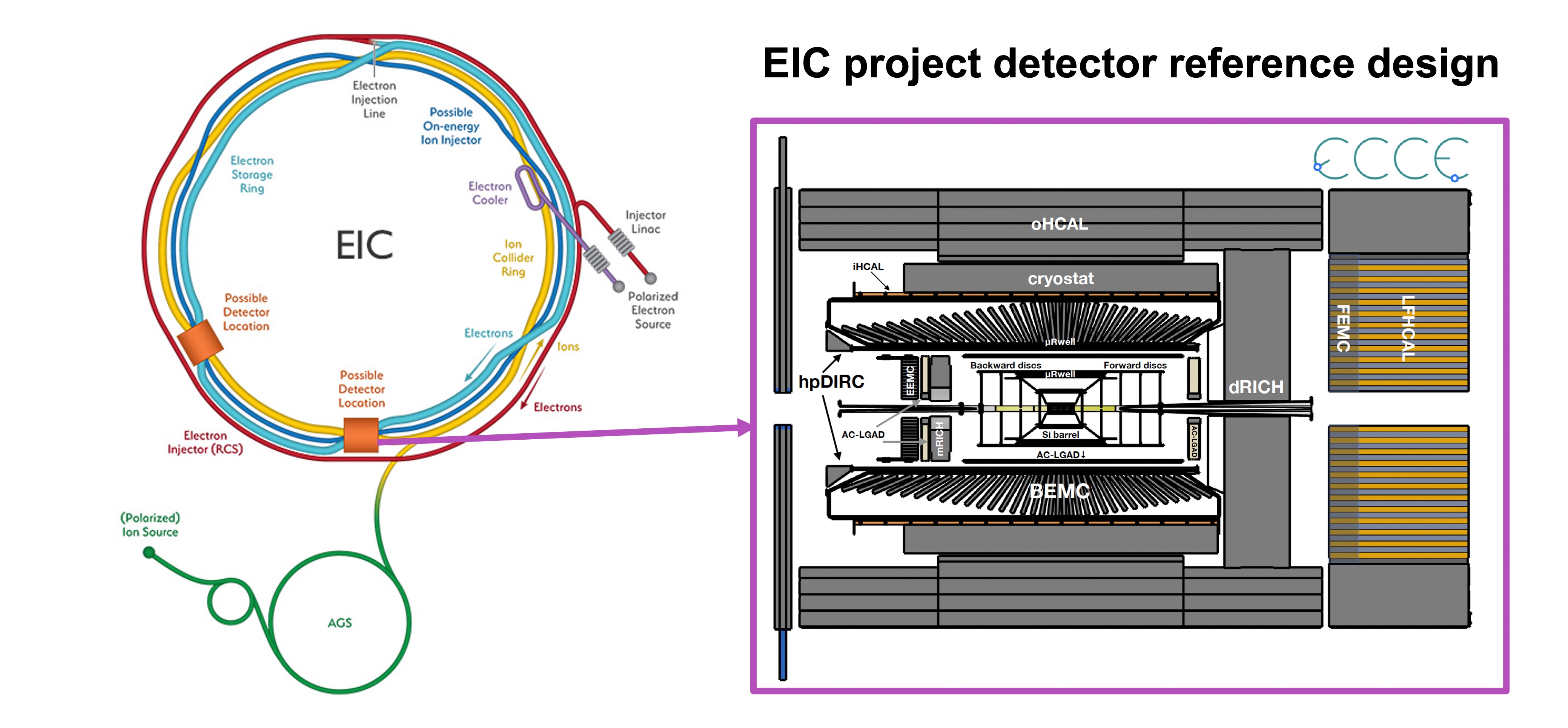}
\caption{The EIC accelerator apparatus (left) has been designed to host the EIC project detector at the 6 o'clock position and potentially a second detector at the 8 o'clock location. The detector layout of the EIC project detector reference design is shown in the right.}
\label{fig:eic_ref}
\end{figure}

The future Electron-Ion Collider (EIC), which will be built at Brookhaven National Laboratory, has received the CD1 approval to define its funding scope in 2021 by the US Department of Energy (DOE). The EIC will operate electron+proton ($e+p$) and electron+nucleus ($e+A$) collisions with the electron beam energy at 2.5-18~GeV and the proton/nucleus beam energy at 41, 100-275~GeV. The Instantaneous luminosities of the EIC are around $10^{33-34}~\rm{cm}^{-2}\rm{s}^{-1}$ and the bunch crossing rate is around 10~ns \cite{eic_YR}. The EIC project detector, which locates at the 6 o'clock location of the accelerator ring, is scheduled to start construction in 2025. The reference design of the EIC project detector \cite{ecce}, which was selected in early 2022, consists of high granularity vertex, tracking, particle identification, electromagnetic calorimeter and hadronic calorimeter subsystems. Figure~\ref{fig:eic_ref} shows the bird-view of the EIC accelerator apparatus (left) and the EIC project detector reference design (right). Following the EIC project detector reference design decision, a new collaboration: ePIC has been formed to lead the EIC project detector design optimization towards the CD3A approval, which is scheduled in 2024.

\section{EIC Project Detector Silicon Tracking Subsystem Design and Performance}
The vertex and tracking subsystem of the EIC project detector reference design utilizes the next generation 65~nm Monolithic Active Pixel Sensor (MAPS) \cite{its3}, Micro-Pattern Gaseous Detector (MPGD) \cite{mpgd} and AC coupled Low Gain Avalanche Diode (AC-LGAD) \cite{ac-lgad} technologies. The EIC project detector design optimization is underway by the ePIC collaboration considering the detector integration needs and performance improvements. Two iterations of geometry optimizations have been performed for the ePIC MAPS vertex and tracking detector design. The updated ePIC MAPS vertex and tracking detector covers a relatively larger active detector area with a larger number of channels than the EIC project detector reference design. The geometry parameters of the current ePIC MAPS vertex and tracking detector are shown in Table~\ref{epic_barrel_trk}, Table~\ref{epic_had_trk} and Table~\ref{epic_ele_trk}. 

The ePIC barrel MAPS vertex and tracking subsystem consists of 3 vertex layers based on bent sensors and 2 sagitta layers of stitched sensor staves. The radii of the MAPS vertex layers vary from 3.6~cm to 12.0~cm and the radii of the MAPS sagitta layers are at 27~cm and 42~cm. The lengths of the barrel MAPS vertex and tracking layers are from 27~cm to 84~cm. This design aims to obtain low material budgets and good tracking momentum and spatial resolutions listed in the EIC yellow report \cite{eic_YR}. The ePIC MAPS hadron endcap detector (geometry parameters listed in Table~\ref{epic_had_trk}) is composed of 5 disks with expanded z coverage from 25 cm to 135 cm. Due to the constraints of the EIC beam pipe dimensions and the allocated detector space limitation, the inner radii of the ePIC MAPS hadron endcap disks vary from 3.67~cm to 7.01~cm and the outer radii vary from 23~cm to 43~cm. Compared to the EIC project detector reference design, the ePIC MAPS electron endcap detector (geometry parameters listed in Table~\ref{epic_ele_trk}) includes one additional disk and the longitudinal and radial parameters of all five disks have been updated to make full usage of the allocated space. The pixel size of the current ePIC MAPS vertex and tracking detector design is assumed at 10 $\mu \rm{m}$, and is open fors further tuning depends on the progress of the sensor design. The material budgets of the ePIC MAPS vertex detector design can reach as low as 0.05$\%~X/X_{0}$ per layer and the material budgets are expected to not exceed 0.55$\%~X/X_{0}$ per layer for the outer most MAPS batter layer.

\begin{table}[tbh]
\centering
\caption{ePIC MAPS barrel vertex and tracking detector geometry}
\label{epic_barrel_trk}
\begin{tabular}{|clc|c|c|c|c|}
\hline
Index  & $r$ (cm) & $z_{\rm min}$ (cm) & $z_{\rm max}$ (cm) & Pixel Pitch ($\mu \rm{m}$) & Material Budget ($X/X_{0}$) \\  \hline
 1  & 3.6 & -13.5 & 13.5  &  10  & 0.05$\%$  \\ 
 2  & 4.8 & -13.5 & 13.5  &  10  & 0.05$\%$ \\
 3  & 12 & -13.5 & 13.5  &  10  & 0.05$\%$ \\ 
 4  & 27 & -27 & 27 & 10  & 0.25$\%$ \\ 
 5  & 42 & -42 & 42 & 10  & 0.55$\%$ \\
\hline
\end{tabular}
\end{table}

\begin{table}[tbh]
\centering
\caption{ePIC MAPS hadron endcap tracking detector geometry}
\label{epic_had_trk}
\begin{tabular}{|clc|c|c|c|c|}
\hline
Index  & $z$ (cm) & $r_{\rm in}$ (cm) & $r_{\rm out}$ (cm) & Pixel Pitch ($\mu \rm{m}$) & Material Budget ($X/X_{0}$) \\  \hline
 1  & 25 & 3.676 & 23  &  10  & 0.24$\%$  \\ 
 2  & 45 & 3.676 & 43  &  10  & 0.24$\%$ \\
 3  & 70 & 3.842 & 43  &  10  & 0.24$\%$ \\ 
 4  & 100 & 5.443 & 43  &  10  & 0.24$\%$ \\ 
 5  & 135 & 7.014 & 43  &  10  & 0.24$\%$ \\
\hline
\end{tabular}
\end{table}

\begin{table}[tbh]
\centering
\caption{ePIC MAPS electron endcap tracking detector geometry}
\label{epic_ele_trk}
\begin{tabular}{|clc|c|c|c|c|}
\hline
Index  & $z$ (cm) & $r_{\rm in}$ (cm) & $r_{\rm out}$ (cm) & Pixel Pitch ($\mu \rm{m}$) & Material Budget ($X/X_{0}$) \\  \hline
 1  & -25 & 3.676 & 23  &  10  & 0.24$\%$  \\ 
 2  & -45 & 3.676 & 43  &  10  & 0.24$\%$ \\
 3  & -65 & 3.676 & 43  &  10  & 0.24$\%$ \\ 
 4  & -90 & 4.006 & 43  &  10  & 0.24$\%$ \\ 
 5  & -115 & 4.635 & 43  &  10  & 0.24$\%$ \\
\hline
\end{tabular}
\end{table}

\begin{table}[tbh]
\centering
\caption{ePIC AC-LGAD outer tracker detector geometry}
\label{epic_aclgad_trk}
\begin{tabular}{|clc|c|c|c|}
\hline
Region  & $z$ locations (cm) & $r$ locations (cm) & pixel size (mm) & Material Budget ($X/X_{0}$) \\  \hline
 Barrel  & -120 to 120 & 64.6  &  $0.5\times1.0$  & $\sim$ 1$\%$  \\ 
h-endcap & 192 & 8.5 to 67 &  $0.5\times0.5$  & $\sim$ 7$\%$ \\
 \hline
\end{tabular}
\end{table}

The current ePIC tracking detector includes the AC-LGAD outer tracker subsystem in the barrel and hadron endcap region as well. The geometry parameters of the AC-LGAD outer tracker are listed in Table~\ref{epic_aclgad_trk}. The pixel size of the barrel AC-LGAD tracker is 0.5~mm by 1.0~mm and the hadron endcap AC-LGAD tracker pixel size is 0.5~mm by 0.5~mm. The barrel AC-LGAD tracker consists of around 2.4 million pixels to cover around 10.9~$\rm{m}^{2}$ active area. The hadron endcap AC-LGAD tracker is composed of around 8.8 million channels with around 2.22~$\rm{m}^{2}$ active area. The barrel AC-LGAD tracker is expected to achieve around 30~$\mu \rm{m}$ spatial resolution in the $r\varphi$ plane and around 30~ps timing resolution. The hadron endcap AC-LGAD tracker is designed to obtain around 30~$\mu \rm{m}$ spatial resolution in the $xy$ plane and around 25~ps timing resolution.

\begin{figure}[!ht]
\centering
\includegraphics[width=0.96\textwidth]{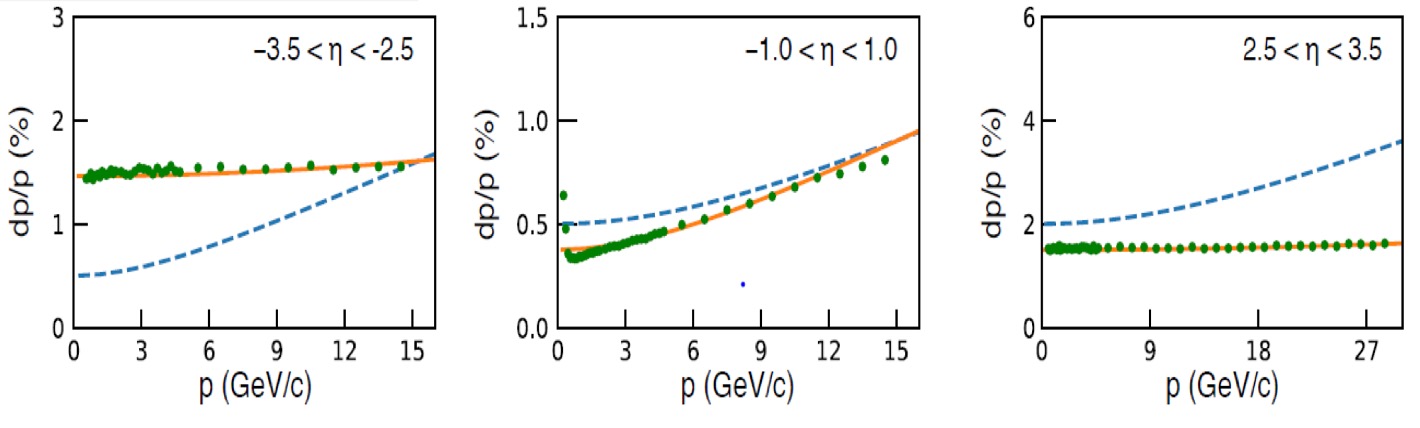}
\includegraphics[width=0.96\textwidth]{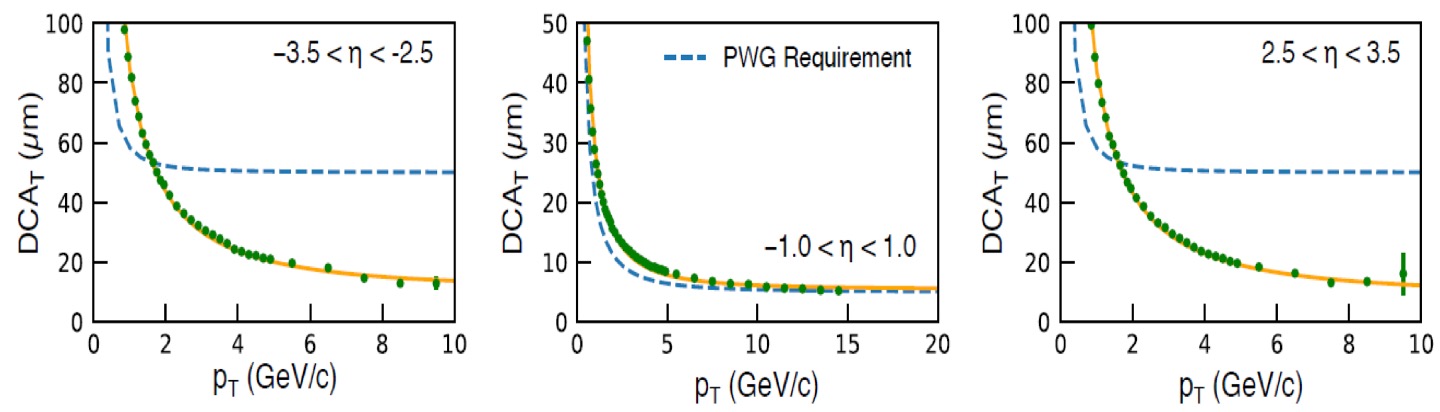}
\caption{Tracking performance of the current ePIC detector design in the pseudorapidity regions of $-3.5 < \eta < -2.5$, $-1.0 < \eta < 1.0$ and $2.5 < \eta < 3.5$. The track momentum dependent momentum resolution (top row) and transverse momentum dependent transverse Distance of Closest Approach ($\rm{DCA}_{\rm{2D}}$) resolution (bottom row) are evaluated with the new 1.7~T ePIC magnet. The EIC yellow report tracking requirements in the respective pseudorapidity regions are highlighted in blue dashed lines.}
\label{epic_trk_perform}
\end{figure}

To evaluate the ePIC detector tracking performance, the active detector volume, the related mechanical structure and the estimated service parts of the current ePIC detector design have been implemented in GEANT4 \cite{geant4} simulation. Figure~\ref{epic_trk_perform} shows the tracking momentum and spatial resolutions of the current ePIC detector design in GEANT4 simulation. With the optimized ePIC detector design, the track momentum dependent momentum resolutions in the pseudorapidity regions of $-1 < \eta < 1$ and $2.5 < \eta < 3.5$ meet the EIC yellow report detector requirements as shown in the top row of Fig.~\ref{epic_trk_perform}. Although the tracking momentum resolution of the current ePIC detector design in the pseudorapidity region of $-3.5 < \eta < -2.5$ is worse than the EIC yellow report detector requirements, further studies to optimize the integrated detector subsystems and reduce the detector material budgets in the electron endcap region are underway. The transverse momentum dependent transverse Distance of Closest Approach ($\rm{DCA}_{\rm{2D}}$) resolution in most pseudorapidity regions meets the EIC yellow report requirements. As the EIC backgrounds such as the beam gas and synchrotron radiation backgrounds, are still under study in simulation, additional tuning of the ePIC tracking detector geometry is expected to provide sufficient number of hits for tracking pattern recognition studies. The ePIC detector tracking performance will be updated accordingly with new simulation samples which will include the EIC backgrounds.

\section{EIC Silicon Detector Technology and R$\&$D Status}
\begin{figure}[!ht]
\centering
\includegraphics[width=0.85\textwidth]{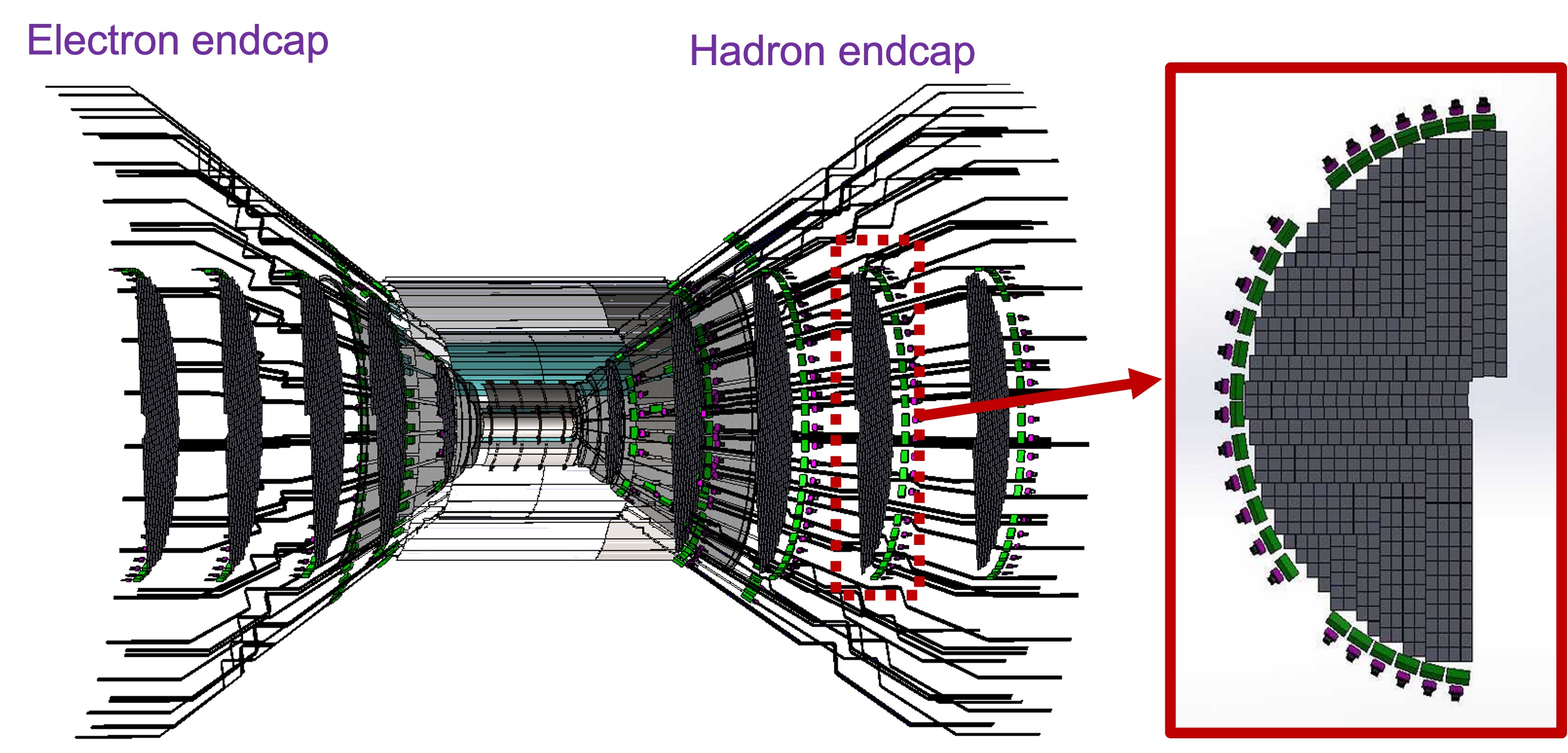}
\caption{The mechanical design of the ePIC MAPS vertex and tracking detector according to the current optimized ePIC detector geometry (details in Table~\ref{epic_barrel_trk}, Table~\ref{epic_had_trk} and Table~\ref{epic_ele_trk}). As highlighted inside the red box, the current MAPS disk layout is based on stitched flat MAPS sensors.}
\label{fig:maps_mech}
\end{figure}

The EIC silicon R$\&$D supported by the EIC project detector program primarily focuses on the EIC MAPS and AC-LGAD sensor developments, vertex and tracking detector mechanical design and readout options. The 65~nm MAPS sensor R$\&$D for the EIC silicon vertex detector aligns well with the ALICE ITS3 developments \cite{its3}. Better than 4~$\mu \rm{m}$ spatial resolution has been achieved by the ALICE ITS3 prototype sensor beam tests with 15~$\mu \rm{m}$ pixel size and $10^{15}~\rm{n}_{eq}\rm{cm}^{-2}$ irradiation dose \cite{its3-d1, its3-d2}. Additional developments towards 0.05$\%~X/X_{0}$ material budgets per layer, on the order of 100~ns timing resolution and low fake-hit rate are ongoing. The EIC flat MAPS prototype sensors for the sagitta layers and endap disks are still under design to obtain similar features of the ALICE ITS3 technology. Figure~\ref{fig:maps_mech} shows the mechanical design of the ePIC MAPS vertex and tracking detector current design (see geometry parameters in Table~\ref{epic_barrel_trk}, Table~\ref{epic_had_trk} and Table~\ref{epic_ele_trk}) with the evaluated support structure, power and data cables and cooling system. Other components such as the readout modules are under developments.

\begin{figure}[!ht]
\centering
\includegraphics[width=0.85\textwidth]{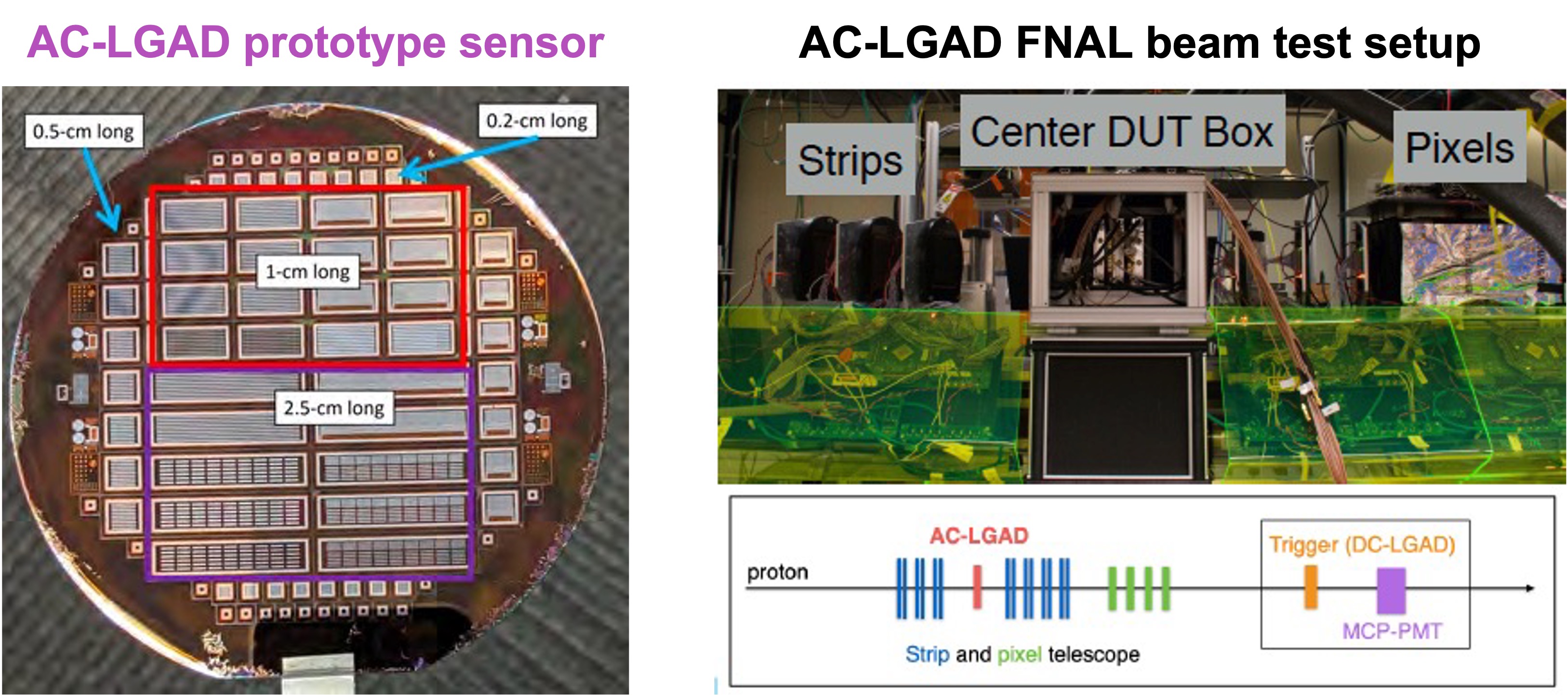}
\caption{EIC AC-LGAD prototype sensor production at BNL and beam test setup at FNAL. The left panel shows the EIC AC-LGAD pixel and strip sensors have been diced on a silicon wafer at BNL. The configuration of AC-LGAD prototype sensor beam tests at the FNAL test beam facility is shown in the right panel.}
\label{fig:aclgad_stat}
\end{figure}

The ePIC AC-LGAD outer tracker will use two types of sensors: 0.5~mm by 0.5~mm pixel sensor and 0.5~mm by 1.0~mm strip sensor. As shown in the left panel of Figure~\ref{fig:aclgad_stat}, new AC-LGAD pixel and strip prototype sensors have been produced at BNL and Hamamatsu Photonics K.K. (HPK) company. The AC-LGAD strip sensors with different strip lengths have been characterized with the 120~GeV proton beam at FNAL (see the setup configuration in the right panel of Figure~\ref{fig:aclgad_stat}). Around 30~$\mu \rm{m}$ spatial resolution and better than 30~ps timing resolution per hit have been achieved from the FNAL AC-LGAD strip sensor beam tests, which meets the EIC AC-LGAD detector design requirement.

\begin{figure}[!ht]
\centering
\includegraphics[width=0.9\textwidth]{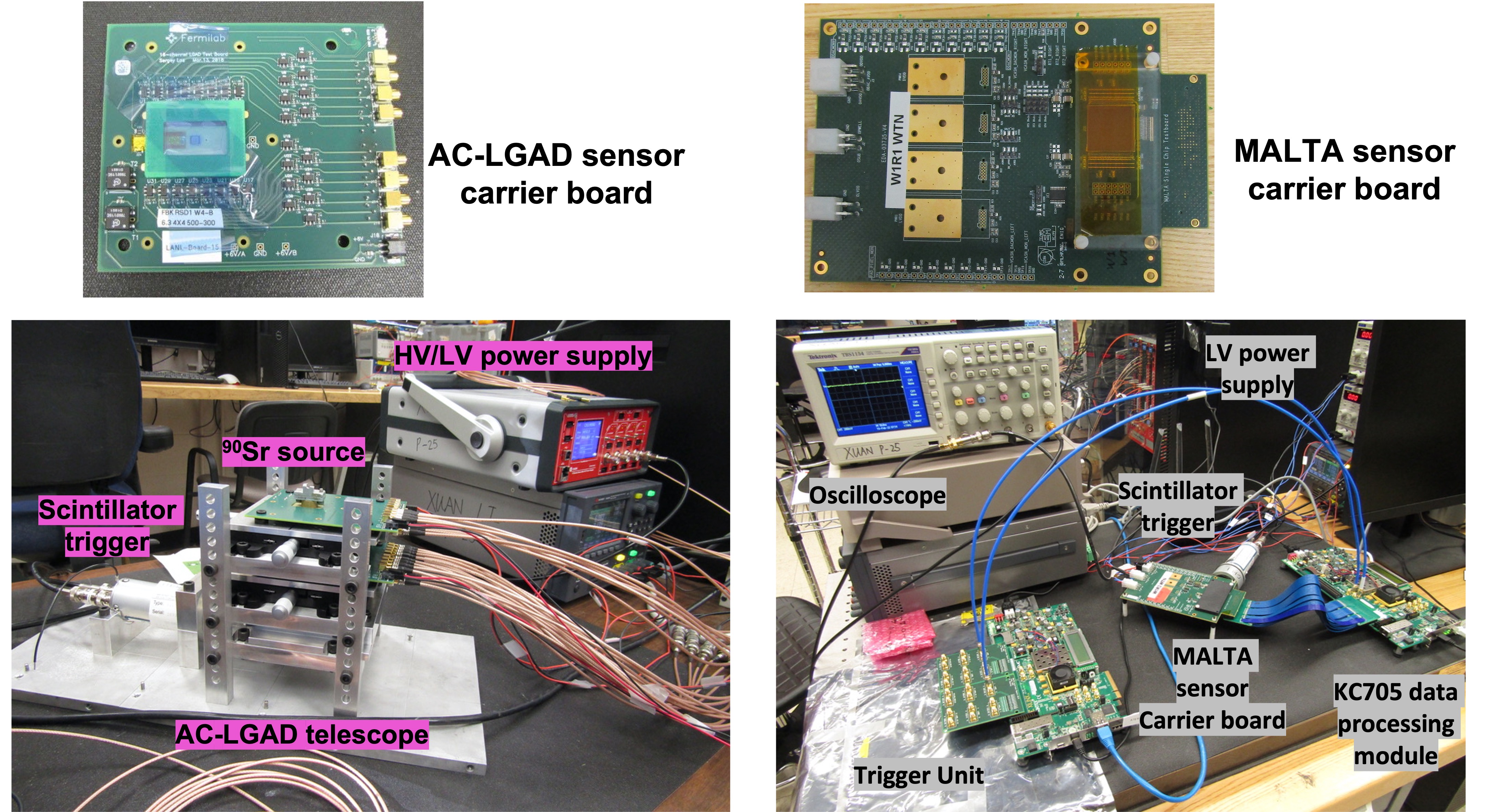}
\caption{The carrier board of the AC-LGAD prototype pixel sensor and the 3-layer AC-LGAD sensor telescope $^{90}\rm{Sr}$ source test configuration is shown in the left panel. The MALTA prototype sensor carrier board and the bench test configuration of a single MALTA sensor is shown in the right panel.}
\label{fig:aclgad_malta_test}
\end{figure}

Other advanced silicon prototype sensors have been studied in parallel for the EIC detectors. Through supports by the Los Alamos National Laboratory (LANL) Laboratory Directed Research $\&$ Development (LDRD), the LANL EIC team has setup the test bench and performed characterizations of LGAD, AC-LGAD and Depleted MAPS (DMAPS) prototype sensors in collaboration with other institutions. Figure~\ref{fig:aclgad_malta_test} shows the bench setup of the 3-layer AC-LGAD pixel sensor telescope $^{90}\rm{Sr}$ source tests and the DMAPS MALTA \cite{malta} sensor bench tests. The pixel size of the AC-LGAD prototype pixel sensor is 500~$\mu \rm{m}$ and the MALTA prototype sensor pixel size is 36.4~$\mu \rm{m}$. The AC-LGAD prototype sensor contains a 4 by 4 pixel array and the MALTA prototype sensor is composed of a 512 by 512 pixel matrix. The AC-LGAD prototype sensor bench test system consists of a scintillator trigger unit to trigger on $^{90}\rm{Sr}$ beta decay electron events, low and high power supplies to operate the sensors and their carrier boards, a 2.5~GHz oscilloscope to evaluate the analog signals from the sensor, daisy-chained 16-channel CAEN 1730s digitizers to perform fast digitization, and a Data Acquisition (DAQ) computer to collect the digitized data. For the AC-LGAD sensor data digitization, one 16-channel CAEN 1730s module is connected to one AC-LGAD sensor carrier board, which is wire-bounded to 16 AC-LGAD pixels. The MALTA prototype sensor test system is composed of a XILINX KC705 module for data processing, a scintillator trigger unit to provide the $^{90}\rm{Sr}$ beta decay electron event trigger, a XILINX KC705 module for parallel trigger synchronization and transmission, and a DAQ computer to provide the MALTA sensor operation, online monitoring and data collection. 

\begin{figure}[!ht]
\centering
\includegraphics[width=0.96\textwidth]{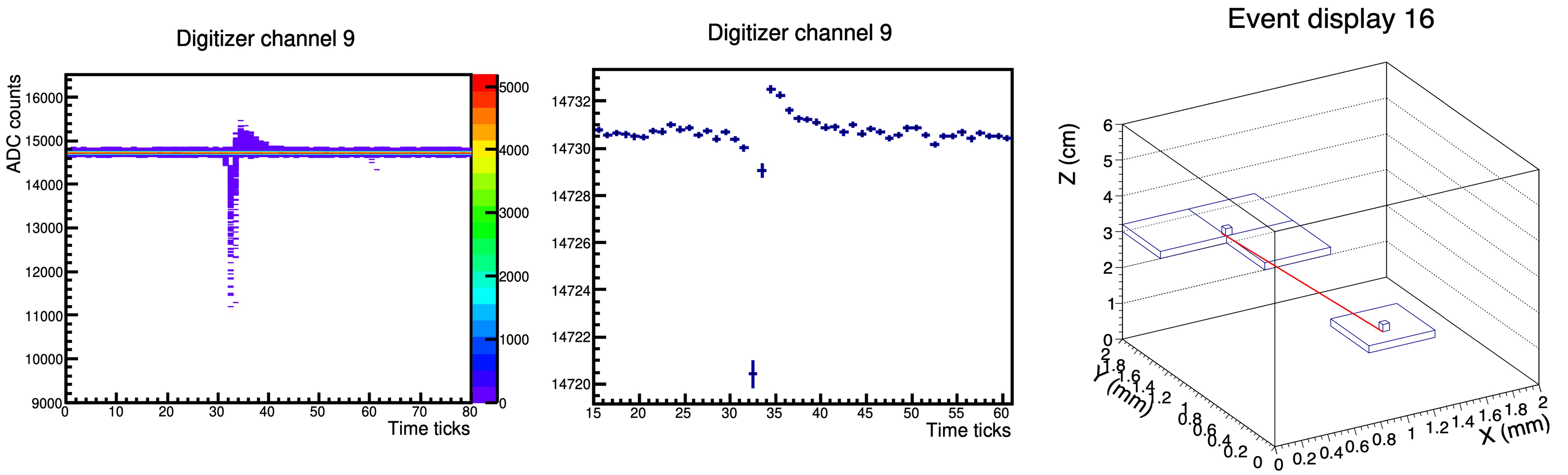}
\caption{The AC-LGAD prototype pixel sensor $^{90}\rm{Sr}$ source test results. The accumulated digitized signals of a AC-LGAD pixel (left) and the average pulse shape of this AC-LGAD pixel (middle) in the $^{90}\rm{Sr}$ source tests. The event display of the reconstructed electron track from clustered hits collected from the three-layer AC-LGAD telescope $^{90}\rm{Sr}$ tests. }
\label{fig:aclgad_sr_res}
\end{figure}

In the AC-LGAD sensor $^{90}\rm{Sr}$ source tests, the amplitude of the AC-LGAD pixel analog signal is around -500~mV and the pulse Full Width Half Maximum (FWHM) is around 1~ns. The left panel of Figure~\ref{fig:aclgad_sr_res} shows the accumulated digitized signals of a single AC-LGAD pixel collected from the $^{90}\rm{Sr}$ source tests and the average pulse shape is shown in the middle panel of Figure~\ref{fig:aclgad_sr_res}. The gain uniformity of all 16 pixels in the AC-LGAD sensor has been verified. A three-layer AC-LGAD prototype sensor telescope has been assembled and tested with the $^{90}\rm{Sr}$ beta decay electrons. For the AC-LGAD telescope tests, synchronized triggers are distributed to the connected CAEN 1730s digitizers and a offline analysis framework has been developed to reconstruct electron tracks from clustered hits located in different AC-LGAD sensors. The right panel of Figure~\ref{fig:aclgad_sr_res} illustrates one event display of a reconstructed electron track (highlighted in red) in the AC-LGAD telescope $^{90}\rm{Sr}$ source tests. The radiation tolerance of the AC-LGAD prototype sensors has been studied in 2022 with the 500~MeV proton beam at the LANL LANSCE facility and the accumulated radiation doses are from $10^{13} - 10^{16}~\rm{n}_{eq} \rm{cm}^{-2}$. Ongoing data analysis will characterize and compare the AC-LGAD sensor performance at different radiation doses. The MALTA sensor bench tests include threshold and noise scan and hit occupancy studies with and without a $^{90}\rm{Sr}$ source \cite{malta_tes}. A new version of the MALTA sensor: MALTA2 \cite{malta2} has been recently produced and characterized with beam tests at CERN. Better than 2~ns timing resolution has been obtained by the MALTA2 sensor. The threshold over noise ratio of the MALTA2 sensor is better than 10 even with $3 \times 10^{15}~\rm{n}_{eq}\rm{cm}^{-2}$ radiation dose.

\section{Summary and Outlook}
The EIC project detector design is evolving rapidly led by the newly formed ePIC collaboration. The MAPS and AC-LGAD based silicon detector subsystems play an essential role in providing precise vertex determination, track reconstruction and particle identification in a broad kinematic coverage for the proposed EIC physics measurements. Good progresses have been achieved for the ePIC MAPS and AC-LGAD detector design optimization and related R$\&$D, which are in good synergy with other major detector development projects, which have similar timeline and schedules, such as the High Luminosity LHC detector upgrade. The current ePIC detector design will deliver an integrated silicon detector that can obtain better than 10~$\mu \rm{m}$ tracking spatial resolution, sub-percent tracking momentum resolution and better than 30~ps timing resolution. Ongoing EIC silicon detector R$\&$D will provide new sensor concepts, updated mechanical design and options for the readout architecture for the EIC detector construction, which is scheduled to start in 2025.

\end{document}